\documentclass[a4paper,11pt]{article}
\pdfoutput=1 % if your are submitting a pdflatex (i.e. if you have
\usepackage{jcappub} % for details on the use of the package, please
\usepackage[T1]{fontenc} % if needed

\newcommand{\be}{\begin{equation}}
\newcommand{\ee}{\end{equation}}
\newcommand{\bea}{\begin{eqnarray}}
\newcommand{\eea}{\end{eqnarray}}
\newcommand{\vp}{\varphi}
\newcommand{\rf}[1]{(\ref{#1})}

\def\K{K{\"a}hler}

\title{\boldmath Random Potentials and Cosmological Attractors}

 \author{Andrei Linde}

\affiliation{Department of Physics and SITP, Stanford University, \\ 
Stanford, California 94305 USA}

\emailAdd{alinde@stanford.edu}
 
\abstract{I show that the problem of realizing inflation in theories with random potentials of a limited number of fields can be solved, and agreement with the observational data can be naturally achieved if at least one of these fields has a non-minimal kinetic term of the type used in the theory of cosmological $\alpha$-attractors.}

\begin{document}
\maketitle
\flushbottom

\section{Introduction}
\label{sec:intro}

There are several different ways of constructing inflationary models. For example, one may try to find the best and simplest models that work and correctly describe observations. Alternatively, one may study the most general theory and check what happens. Since the ideas of what is best, simplest and most general are gradually changing under pressure of observational data and theoretical developments, these two complementary approaches often inform and influence each other.

Both of these approaches are based on the main principle of chaotic inflation  \cite{Linde:1983gd}: One should try to find a theory which may give rise to inflation, and then examine whether this may happen under natural initial conditions, without assuming, for example, that the universe initially must be absolutely homogeneous and isotropic \cite{Starobinsky:1980te}, or that the initial position of the inflaton field must be determined by high-temperature phase transitions  \cite{Guth:1980zm,Linde:1981mu}.

The basic single-field model of chaotic inflation   \cite{Linde:1983gd}, has the Lagrangian
 \be
 {1\over \sqrt{-g}} \mathcal{L} = {1\over 2}   R - {1\over 2} (\partial_{\mu} \phi)^2  - V(\phi)   \,  .
\label{quadratic}
\ee
Recent observational data \cite{Planck:2015xua} disfavor the model with the simplest quadratic potential $V(\phi) = {1\over 2}{m^2} \phi^2$. One can easily remedy this problem by considering slightly more general models $V(\phi) = {1\over 2}{m^2} \phi^2 (1 + a \phi + b\phi^{2})$ or $ {1\over 2}{m^2} \phi^2 (1 + a \phi + b\phi^{2})^{2}$. Then by tuning 3 parameters $m$, $a$, and $b$ one can account for any or almost any values of the 3 observational parameters $A_{s}$, $n_{s}$ and $r$, describing the amplitude of the scalar perturbations, the slope of the spectrum and the tensor to scalar ratio \cite{Destri:2007pv,Nakayama:2013jka,Kallosh:2014xwa}. However, the most natural and economical description of the Planck data, which requires only one parameter responsible for  the amplitude of scalar perturbations $A_{s}$, can be achieved using the models with plateau potentials.

The first example of a model with plateau potentials was given in  \cite{Goncharov:1983mw}, in the supergravity context. It describes a potential exponentially rapidly approaching a  positive constant for super-Planckian values of the inflaton field. Later on, it was realized that the Starobinsky model \cite{Starobinsky:1980te} can be reformulated in terms of a theory with a similar potential \cite{whitt}, and then the Higgs inflation model with a similar potential was developed    \cite{Salopek:1988qh,Bezrukov:2007ep}. These models are very different, but nevertheless they lead to nearly identical predictions, providing the best fit to the latest Planck data  \cite{Planck:2015xua}. This puzzling fact was explained only recently, when these models have been embedded into a broad class of inflationary models, cosmological attractors   \cite{Kallosh:2013wya,Kallosh:2013xya,Kallosh:2013hoa,Ferrara:2013rsa,Kallosh:2013maa,Kallosh:2013daa,Kallosh:2013tua,Kallosh:2013yoa,Galante:2014ifa,Kallosh:2015zsa,Kallosh:2016gqp}, which lead to very similar  cosmological predictions.

In parallel to these developments triggered by observational data, cosmologists were trying to investigate theories of many scalar fields $\phi_{i}$ with random potentials $V(\phi_{i})$,  see e.g. \cite{Aazami:2005jf,Frazer:2011tg,Battefeld:2012qx,Marsh:2013qca,Bachlechner:2014rqa,Dias:2016slx,Freivogel:2016kxc,Masoumi:2016eag,Easther:2016ire}. Such potentials may play the role  of a toy model for the string theory landscape \cite{Bousso:2000xa,Kachru:2003aw,Susskind:2003kw,Douglas:2003um}. Some authors assumed that random  potentials should strongly vary on the sub-Planckian scale $\Delta\phi_{i}\lesssim 1$, which may make inflation problematic unless one takes into account possible existence of flat directions. 
On the other hand, if the potentials change slowly on the super-Planckian scale $\Delta\phi_{i}\gtrsim 1$, they can support slow-roll inflation, but as we already seen in the single-field case, observational data may require the potential to have some special structure instead of being random. 

 In particular, if the observational data will continue to  push us towards theories with plateau potentials, we will need to understand how such potentials may ``randomly'' appear.  The goal of this paper is to explain how one can get plateau potentials starting from models with random potentials, if at least one of the fields has a non-minimal kinetic term of the type used in the theory of cosmological $\alpha$-attractors.

\section{\boldmath Single field $\alpha$-attractors} 

There are many different versions of the theory of cosmological attractors, with different origin and motivation, but their main features can be explained by making a simple modification of the kinetic term in the model (\ref{quadratic}):
 \be
 {1\over \sqrt{-g}} \mathcal{L} = {1\over 2}   R - {1\over 2} {(\partial_{\mu} \phi)^2\over (1-{\phi^{2}\over 6\alpha})^{2}}  -  V(\phi)   \,  .
\label{alpha}
\ee
This model provides the simplest example of the $\alpha$-attractor \cite{Kallosh:2013yoa}. In the limit $\alpha \to \infty$, this model coincides with \rf{quadratic}. However, for any finite $\alpha$, the absolute value of the field $\phi$ cannot become greater than $\sqrt{6\alpha}$, because the kinetic terms blows up there. But in fact, the field can never reach this boundary. Indeed, one can describe the same theory in terms of a canonically normalized inflaton field $\vp$ such that
\be\label{tanh} 
\phi = \sqrt {6 \alpha}\, \tanh{\varphi\over\sqrt {6 \alpha}} \ .
\ee
In terms of the canonical variables, the boundary $|\phi| =\sqrt{6\alpha}$ moves to infinity, and the theory becomes
 \be
 {1\over \sqrt{-g}} \mathcal{L} = { R\over 2}   -  {(\partial_{\mu}\varphi)^{2} \over 2}  - V\big(\sqrt {6 \alpha}\, \tanh{\varphi\over\sqrt {6 \alpha}}\big)   \,  .
\label{cosmoqq}\ee

In the vicinity of the boundary $\phi=\sqrt {6 \alpha}$, the relation \rf{tanh} between the original field variable $\phi$ and the canonically normalized inflaton field $\vp$ is given by
\be\label{tanh2} 
\phi = \sqrt {6 \alpha}\, \left(1 - 2 e^{-\sqrt{2\over 3\alpha} \varphi }\right)\ ,
\ee
up to the higher order terms $O\bigl(e^{-2\sqrt{2\over 3\alpha} \varphi }\bigr) $. At $\vp \gg \sqrt \alpha$, these  terms are exponentially small as compared to the terms $\sim  e^{-\sqrt{2\over 3\alpha} \varphi }$, and the potential acquires the following asymptotic form:
\be\label{plateau}
V(\vp) = V_{0} - 2  \sqrt{6\alpha}\, V'_{0}\ e^{-\sqrt{2\over 3\alpha} \varphi } \ .
\ee
Note that the constant $2  \sqrt{6\alpha}\, V'_{0}$ in this expression can be absorbed into a redefinition (shift) of the field $\vp$. This implies that if inflation occurs at large $ \vp \gg \sqrt{\alpha}$, all inflationary predictions in this class of models  are determined only by the value of the potential $V_{0}$ at the boundary and the constant $\alpha$. For $\alpha = O(1)$, the predictions for the amplitude $A_{s}$ requires
\be
{V_{0}\over  \alpha} \sim 10^{{-10}} \ .
\ee
The prediction for  $n_{s}$ and $r$ are 
\be\label{nsr}
 1 -n_{s}  \approx {2\over N}\, , \qquad r  \approx  {12\alpha \over N^{2} } \ .
 \ee
For the number of e-foldings $N \sim 55$, these predictions match the observational data without any fine-tuning for all of these models. In the future, when we will know $n_{s}$ and $r$ with better accuracy, subtle distinctions between various models, including the shape of the potential at small $\varphi$ and the mechanism of reheating, will become important, which will help us to distinguish between different models, see e.g.  \cite{Ueno:2016dim,Eshaghi:2016kne}. 
   
\begin{figure}[tbp]
\centering % \begin{center}/\end{center} takes some additional vertical space
\includegraphics[width=.4\textwidth]{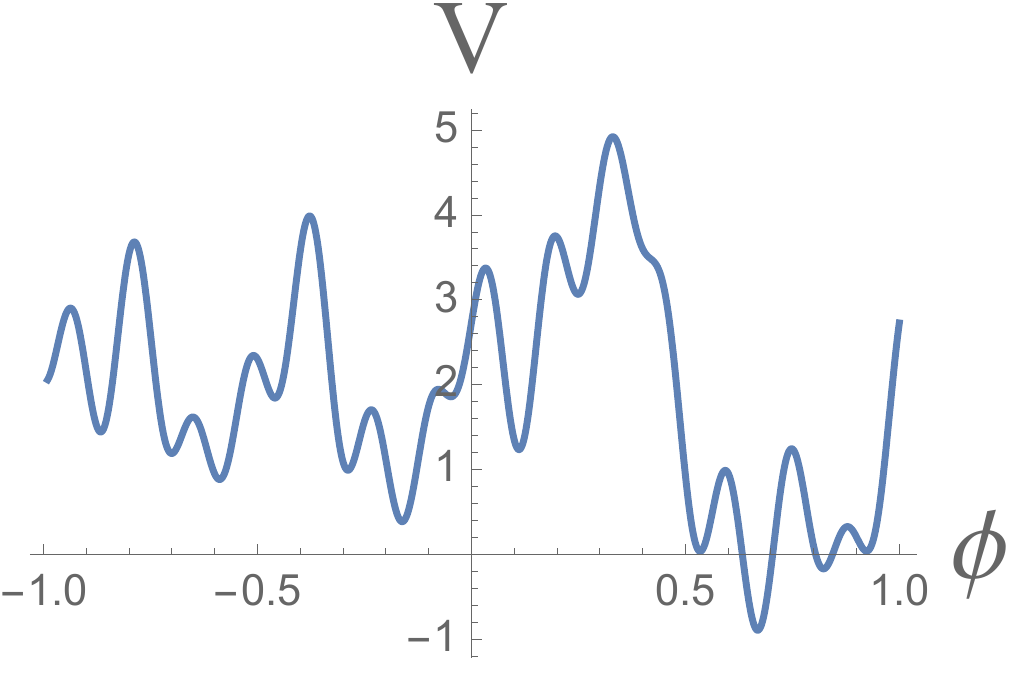}
% "\includegraphics" is very powerful; the graphicx package is already loaded
\caption{\label{fig:1} Random potential $V(\phi)$ of a single field $\phi$ in the theory  with $6\alpha = 1$.}
\end{figure}

To develop an intuitive understanding of the properties of $\alpha$-attractors, we will follow  \cite{Kallosh:2013hoa} and illustrate what is going on there using as an example a random potential of the field $\phi$ in the theory with $6\alpha = 1$, as shown in Fig. \ref{fig:1}. In that case, the theory is defined for $-1 < \phi < 1$. 

\begin{figure}[tbp]
\centering % \begin{center}/\end{center} takes some additional vertical space
\includegraphics[width=0.9\textwidth]{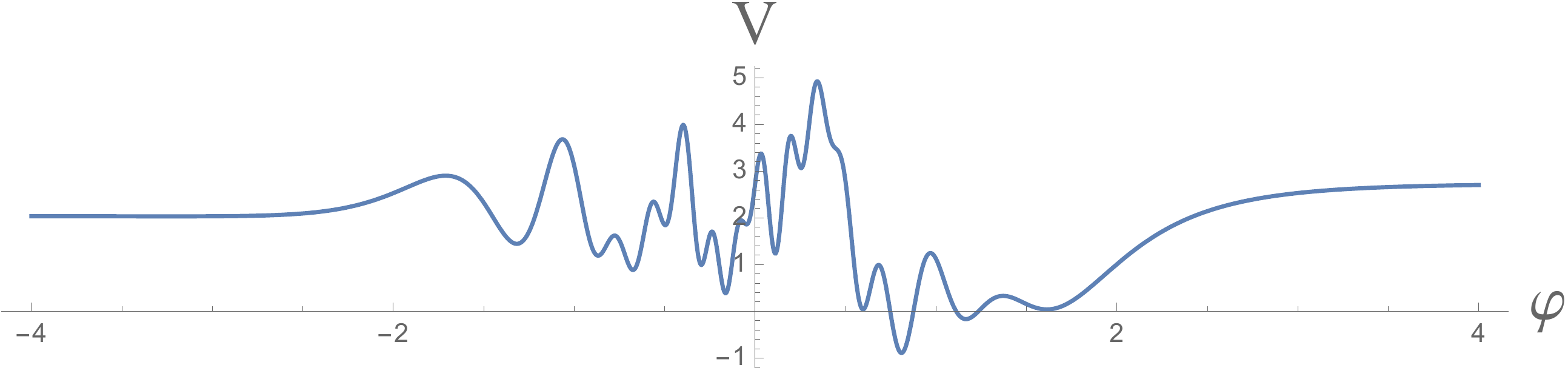}
% "\includegraphics" is very powerful; the graphicx package is already loaded
\caption{\label{fig:2} The same potential in terms of the canonical inflaton field $\varphi$ \rf{tanh}. As we see, the shape of the potential at $\phi \ll 1$ practically did not change. Meanwhile the vicinity of the boundary of the moduli space at $|\phi| = 1$ is infinitely stretched. The height of the potential $V(\vp)$ at $\vp \to \pm \infty$ coincides with $V(\phi)$ at the boundaries of the moduli space $\phi = \pm 1$. }
\end{figure}

If the field $\phi$ were canonically normalized as in \rf{quadratic}, inflation in the theory with the potential shown in Fig. \ref{fig:1} would be impossible. In the theory \rf{alpha}, \rf{cosmoqq} the situation changes dramatically because of the stretching of the potential in terms of the canonically normalized field $\vp$, as shown in Fig. \ref{fig:2}. This stretching does for the inflation potential the same as what inflation does for the universe: It makes the potential exponentially flat. Then inflation becomes possible, and it makes the universe exponentially large, flat and homogeneous. 

Now it is time to issue some warning: Random potentials would not place the inflaton field to the Minkowski or dS vacuum with the tiny cosmological constant $V = 10^{{-120}}$. In this single-field inflation model I do not make any attempts to address the cosmological constant problem, I am just assuming that it is small in one of the string theory vacua. To reflect this assumption, I appropriately uplifted the otherwise random potential. Fortunately, due to the magic of $\alpha$ attractors, this uplifting does not change the  predictions for $n_{s}$ and $r$.

\section{Two-field  $\alpha$-attractors}

Now we will generalize these results for the theory of two field inflation, $\phi$ and $\sigma$, with the Lagrangian
 \be
 {1\over \sqrt{-g}} \mathcal{L} = { R\over 2}   -  {(\partial_{\mu} \phi)^2\over 2(1-{\phi^{2}\over 6\alpha})^{2}} -  {(\partial_{\mu} \sigma)^2\over 2}  - V(\phi,\sigma) .
\label{cosmo2}\ee
In terms of canonical fields $\vp$ with the kinetic term  $ {(\partial_{\mu} \vp)^2\over 2}$, the  potential is 
 \be
 V(\varphi,\sigma) =  V(\sqrt {6 \alpha}\, \tanh{\vp\over\sqrt {6 \alpha}},\sigma).
\label{cosmo3}\ee
During inflation at  $|\varphi | \gg \sqrt\alpha$, one can use the asymptotic equation
 \be
 V(\vp,\sigma)_{|\vp|\gg  \sqrt{6\alpha}}\ \approx \ V(\phi,\sigma)_{\phi = \pm \sqrt {6 \alpha}}\ ,
\label{cosmo3a}\ee
which means that asymptotically $V(\varphi,\sigma)$ is given by the values of the original potential $V(\phi,\sigma)$ at the boundaries of the moduli space. The same is true for the curvature of the potential in the $\sigma$ direction, i.e. for the effective mass squared of the field $\sigma$, which asymptotically approaches a constant value \cite{Kallosh:2016gqp}
 \be
 V_{\sigma,\sigma}(\vp,\sigma)_{|\vp|\gg  \sqrt{6\alpha}}\ \approx \ V_{\sigma,\sigma}(\phi,\sigma) _{\phi = \pm \sqrt {6 \alpha}} \ .
\label{cosmo3b}\ee

\begin{figure}[tbp]
\centering % \begin{center}/\end{center} takes some additional vertical space
\includegraphics[width=0.6\textwidth]{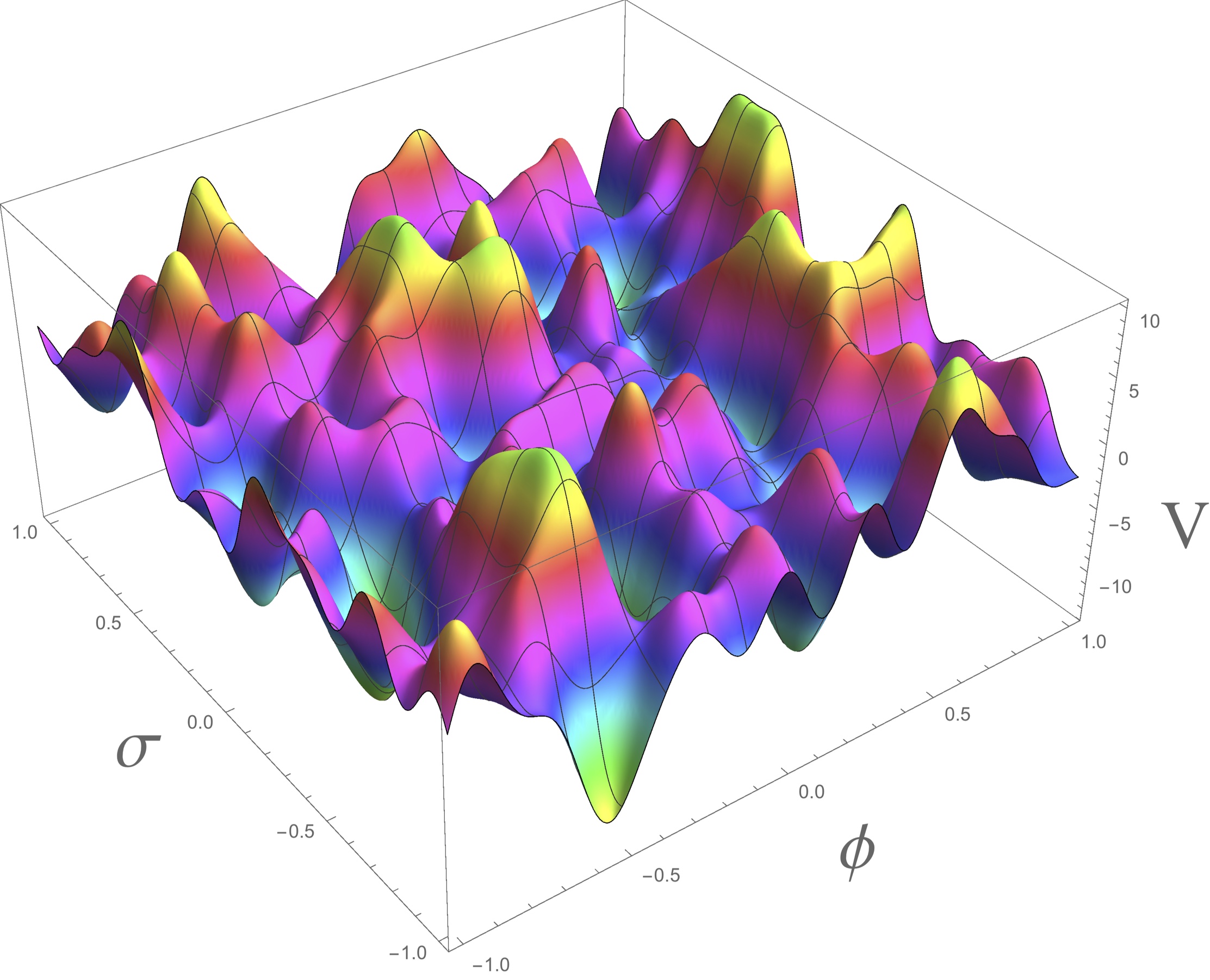}
% "\includegraphics" is very powerful; the graphicx package is already loaded
\caption{\label{fig:3} Randomly generated potential $V(\phi,\sigma)$  in the theory  with $6\alpha = 1$.}
\end{figure}

To illustrate the implications of this result, we will consider again the case $6\alpha = 1$ and generate a random potential $V(\phi,\sigma)$ of the original fields $\phi$ and $\sigma$ in the Planck size box $1 < \phi,\sigma < 1$, see Fig. \ref{fig:3}. Just as in the single field case, the potential $V(\phi,\sigma)$ shown in Fig. \ref{fig:3} is very steep, so it would not support slow roll inflation if both fields were canonically normalized. (We could always generate a smooth potential with the super-Planckian field variations, but we want to analyze the most difficult case when the potential $V(\phi,\sigma)$ is very steep.)

\begin{figure}[tbp]
\centering % \begin{center}/\end{center} takes some additional vertical space
\includegraphics[width=0.9\textwidth]{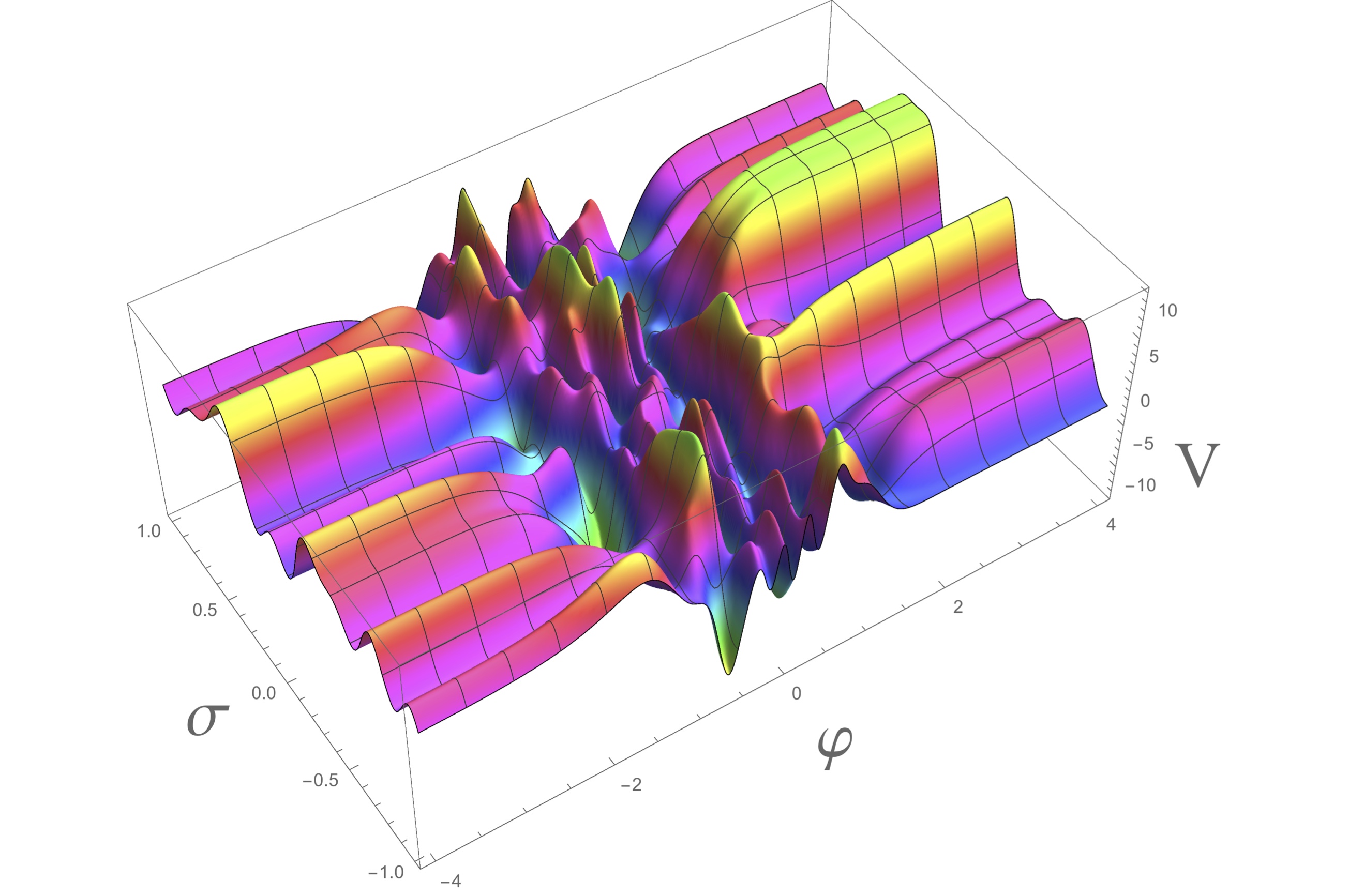}
% "\includegraphics" is very powerful; the graphicx package is already loaded
\caption{\label{fig:4} The same potential in terms of  the field $\sigma$ and the canonical inflaton field $\varphi$. One can see, in particular, that the tops of the three  hills in the upper corner of the previous figure become converted into three infinitely long ridges, and the two minima separating them at $\phi = 1$ in Fig. \ref{fig:3} become two inflationary valleys. }
\end{figure}

The situation looks dramatically different if one plots the same potential in terms of the canonically normalized inflaton field $\vp$, see Fig. \ref{fig:4}. Just as in the one-field case, the part of the potential $V(\phi,\sigma)$ close to the boundaries $\phi = \pm 1$ becomes infinitely stretched, see Fig. \ref{fig:4}. In particular, the tops of the three  hills in the upper corner of Fig. \ref{fig:3} become converted into three infinitely long ridges in Fig. \ref{fig:4}, and the two minima separating them at $\phi = 1$ in Fig. \ref{fig:3} become two infinitely long inflationary valleys. As we already mentioned, the mass squared of the field $\sigma$ along these valleys asymptotically approaches a constant value, which can be calculated directly at $\phi \to 1$ \cite{Kallosh:2016gqp}. Thus if the valley begins at a point corresponding to a minimum of $V(\phi,\sigma)$ with respect to $\sigma$ at the boundary $\phi = \pm 1$, then it describes an  infinitely long inflationary valley, which remains stable until the field $\vp$ becomes sufficiently small.

There are several things that could go wrong about these models. First of all, any of these flat directions can be an AdS valley with a negative value of the potential. Moreover, an inflationary  valley may eventually brings the fields towards an AdS vacuum state with a negative vacuum energy, or to a  dS vacuum state with a very large positive vacuum energy. However, one can adjust the value of the cosmological constant in the context of the string theory landscape, as discussed in the previous section; see also a discussion below. The second issue is apparent from  Figs. \ref{fig:1} and \ref{fig:2}: The potential $V(\vp,\sigma)$ is inflationary and does not push the field infinitely far away if $V(\phi,\sigma)$ grows towards the boundary at $\phi = \pm 1$. This should happen in 50\% of all cases in the random landscape, so this is not a real issue.

\begin{figure}[tbp]
\centering % \begin{center}/\end{center} takes some additional vertical space
\includegraphics[width=0.77\textwidth]{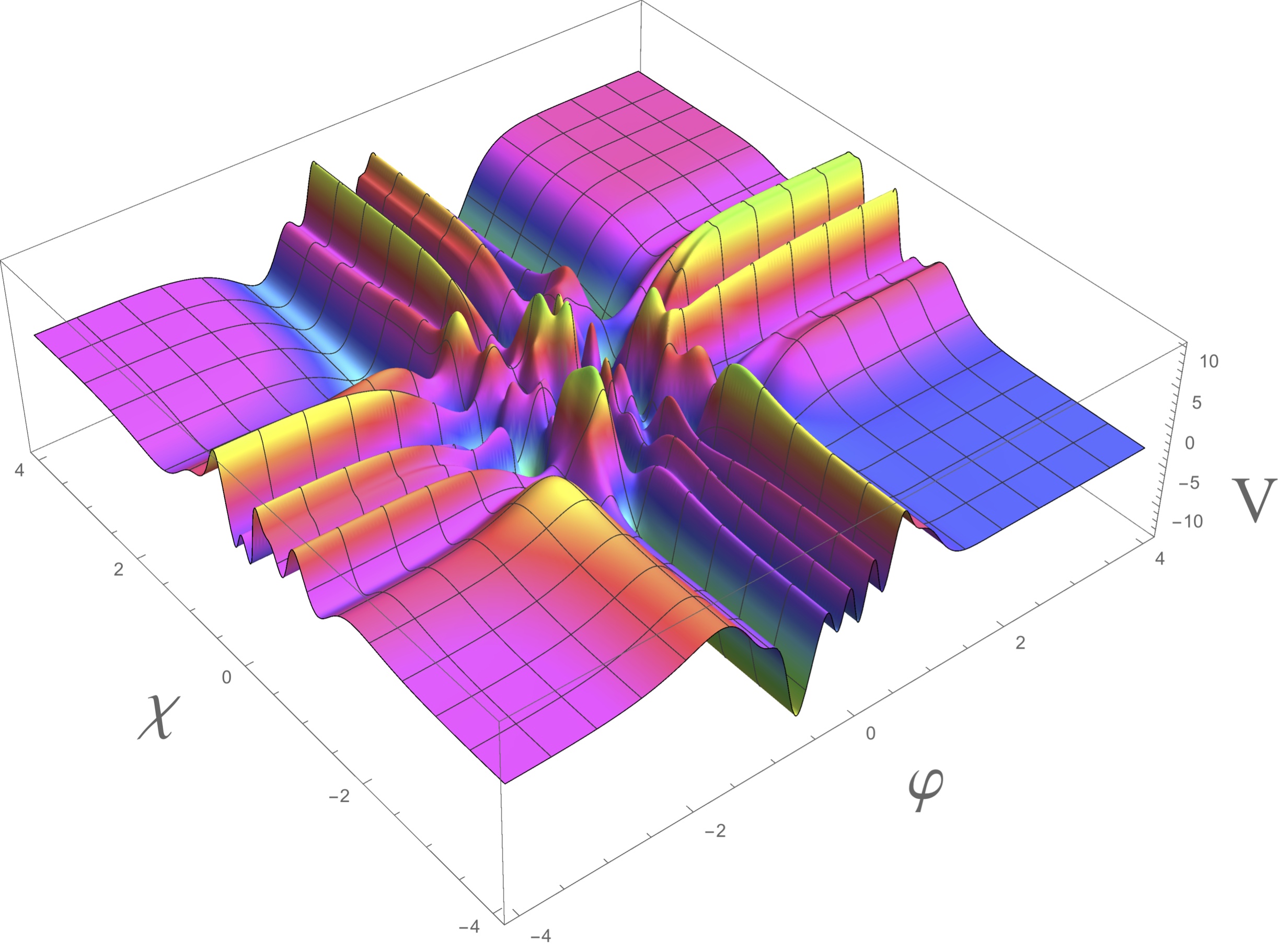}
% "\includegraphics" is very powerful; the graphicx package is already loaded
\caption{\label{fig:5} The random potential shown in Fig. \ref{fig:3} acquires two families of inflationary flat directions in terms of the canonical  fields $\varphi$ and $\chi$ in the context of the double-attractor model \rf{cosmo22}, \rf{cosmo32}.  The potential is shown for particular values $ \alpha =  \beta = 1/6$.}
\end{figure}

Once can also make an additional generalization and modify the kinetic term of the second field as well:
\be
 {1\over \sqrt{-g}} \mathcal{L} = { R\over 2}   -  {(\partial_{\mu} \phi)^2\over 2(1-{\phi^{2}\over 6\alpha})^{2}} -  {(\partial_{\mu} \sigma)^2\over  2(1-{\sigma^{2}\over 6\beta})^{2}}  - V(\phi,\sigma) .
\label{cosmo22}\ee
Then one can make a field redefinition and instead of the fields $\phi$ and $\sigma$ consider the potential in terms of the canonically normalized fields $\vp$ and $\chi$ with kinetic terms  $ {(\partial_{\mu} \vp)^2\over 2}$, and   $ {(\partial_{\mu} \chi)^2\over 2}$.  The  potential in terms of the canonical variables $\vp$ and $\chi$ becomes 
 \be
 V(\varphi,\chi) =  V(\sqrt {6 \alpha}\, \tanh{\vp\over\sqrt {6 \alpha}},\sqrt {6 \beta}\, \tanh{\chi\over\sqrt {6 \beta}}).
\label{cosmo32}\ee
As a result, the random potential shown in Fig. \ref{fig:3}  acquires additional set of flat directions shown in Fig.   \ref{fig:5}.

Note that now we have two sets of inflationary attractors. If inflation occurs when the field  $\vp$ rolls along the inflationary valleys in the $\vp$ direction, the predictions of the theory for $\alpha \lesssim  O(1)$ are 
\be
 1 -n_{s}  \approx {2\over N}\, , \qquad r  \approx  {12\alpha \over N^{2} } \ .
\label{nsra} \ee
However, if  inflation occurs when the field  $\chi$ rolls along the inflationary valleys in the $\chi$ direction, the predictions of the theory for $\beta \lesssim  O(1)$ are
\be
 1 -n_{s}  \approx {2\over N}\, , \qquad r  \approx  {12\beta \over N^{2} } \ .
\label{nsrb} \ee

\section{\boldmath Multifield $\alpha$-attractors}
The results of the previous two sections can be trivially generalized to the case with many different canonically normalized fields $\sigma_{i}$,  $i = 1,...,K$, with the Lagrangian
 \be
 {1\over \sqrt{-g}} \mathcal{L} = {1\over 2}   R - {1\over 2} {(\partial_{\mu} \phi)^2\over (1-{\phi^{2}\over 6\alpha})^{2}}  -  {(\partial_{\mu} \sigma_{i})^2\over  2} -  V(\phi,\sigma_{1},... \sigma_{K})   \,  .
\label{alphaK}
\ee
 Just as in the previous section, one should consider two boundaries, at $\phi = \pm \sqrt{6\alpha}$, and find all minima with respect to all fields $\sigma_{i}$ at each boundary. If there is at least one such minimum, and the potential grows when the field $\phi$ approaches the boundary in the vicinity of this minimum, then the minimum corresponds to the beginning of an infinitely long and stable inflationary valley in terms of the canonically normalized inflaton field $\vp$. 
 
 This multi-valley structure of the potentials in the theory of cosmological attractors appears not only in the theory of $\alpha$-attractors, but also in the theory with non-minimal coupling of the fields to gravity, and in multi-field conformal cosmological attractors; see in particular Figs. 9 and 10 in  \cite{Kallosh:2013daa}.

How many such minima - and such inflationary trajectories - can we find in the random multifield potentials? The answer depends on our assumptions of the statistics of the minima and maxima in the randomly generated potential. In this paper, I described the theory of one or two fields. I performed an investigation of a theory of three fields and found very similar results. Cosmological attractors involving extremely large number of fields \cite{Aazami:2005jf,Frazer:2011tg,Battefeld:2012qx,Marsh:2013qca,Bachlechner:2014rqa,Dias:2016slx,Freivogel:2016kxc,Masoumi:2016eag,Easther:2016ire} should be a subject of a separate investigation. 

\parskip 5pt  

\section{Discussion}
~~~~~~~~Investigation of random potentials in \cite{Aazami:2005jf,Frazer:2011tg,Battefeld:2012qx,Marsh:2013qca,Bachlechner:2014rqa,Dias:2016slx,Freivogel:2016kxc,Masoumi:2016eag,Easther:2016ire} was motivated in part by the string theory landscape \cite{Bousso:2000xa,Kachru:2003aw,Susskind:2003kw,Douglas:2003um}, which has its own distinguishing features, many of which are not fully explored. 

Some of these features appear already at the supergravity level. In particular, most flexible versions of chaotic inflation in $N = 1$ supergravity involve two superfields, the inflaton $\Phi$ and the stabilizer $S$, with the superpotentials $S f(\Phi)$ and \K\ potentials vanishing in the inflaton direction \cite{Kawasaki:2000yn,Kallosh:2010ug}. They can produce random but {\it positively definite} potentials $|f(\Phi)|^{2}$, which typically have supersymmetric Minkowski vacua where $f(\Phi)$ vanishes. One can introduce SUSY breaking and a positive cosmological constant. The simplest way to do it involves nonlinear realization of supersymmetry and nilpotent fields, see e.g. \cite{Ferrara:2014kva,Kallosh:2014via,Dall'Agata:2014oka,Ferrara:2015tyn,Carrasco:2015iij,Dall'Agata:2016yof,Linde:2016bcz}.

Many models of cosmological attractors have been implemented in supergravity. Once it is done, one can introduce interaction of the fields $S$ and $\Phi$ with matter fields, analogous to the field $\sigma$ discussed in this paper. Then one can formulate the conditions that are sufficient for stability and absence of tachyons for all matter fields  \cite{Kallosh:2016ndd}. 

There are some classes of supergravity models where the stability conditions formulated in   \cite{Kallosh:2016ndd}  are not satisfied. However, in the models describing cosmological attractors, the inflationary trajectory usually can be stabilized. Indeed,  due to the nature of the cosmological attractors discussed above, it is sufficient to achieve stabilization at a single point corresponding to the boundary of the moduli space \cite{Kallosh:2016sej}. Tachyonic instabilities may appear in these models after inflation, but thanks to the stabilizing nature of supersymmetry, in some models these instabilities appear to be a harmless part of the tachyonic preheating, eventually bringing all fields to a stable Minkowski or dS vacuum \cite{Kallosh:2016sej}. 

There were many attempts to study stability of KKLT vacua in string theory in the context of supergravity. Recent progress in this direction was achieved due to investigation of nonlinear representations of supersymmetry and the theory of nilpotent fields \cite{Ferrara:2014kva,Kallosh:2014via,Dall'Agata:2014oka,Ferrara:2015tyn,Carrasco:2015iij,Dall'Agata:2016yof,Linde:2016bcz}, combined with direct investigation in the string theory context \cite{Ferrara:2014kva}. Therefore we doubt that investigation of inflation in random multifield potentials, without taking into account the latest developments in supergravity and string theory, can fully represent the real situation. Nevertheless, investigation of such theories may be quite instructive.

In this paper we studied random potential of a limited number of scalar fields.  Our main goal was to show that if at least one of the scalar fields has a non-minimal kinetic term of the type used in the theory of cosmological $\alpha$-attractors, random multifield potentials acquire  many flat directions. If these flat directions correspond to stable inflationary trajectories, the observational predictions of such theories, instead of being random, coincide with the predictions of cosmological attractors \rf{nsr}, providing good fit to the latest observational data \cite{Planck:2015xua}.

\acknowledgments
I am grateful to   R. Kallosh, M.~C.~D.~Marsh, D. Roest and T. Wrase  for  enlightening  discussions.   This work  is supported by the SITP,    by the NSF Grant PHY-1316699, and by the Templeton foundation grant `Inflation, the Multiverse, and Holography.'

\end{document}